\documentclass[aps,prl,amsmath,amssymb,amsfonts,reprint,a4paper,superscriptaddress]{revtex4-1}
\usepackage{newtxtext,newtxmath}
\usepackage[T1]{fontenc}
\usepackage[utf8]{inputenx}
\usepackage{graphicx}
\usepackage{subfig}
\usepackage{color}
\usepackage[dvipsnames]{xcolor}
\usepackage[format=plain,justification=raggedright,singlelinecheck=false]{caption}
\usepackage{soul} 
\usepackage[textwidth=17.5cm,textheight=23.5cm,verbose,pdftex]{geometry}
\usepackage[pdftex]{hyperref}
\usepackage{booktabs} 

\hypersetup{pdfauthor={Dietmar Czubak}, bookmarksnumbered=true, pdftitle={Electronic and magnetic properties of $\alpha$-FeGe$_2$ films embedded in vertical spin valve devices}, colorlinks, citecolor=blue, linkcolor=blue, urlcolor=blue}

\begin{document}

\title{Electronic and magnetic properties of $\alpha$-FeGe$_2$ films embedded in vertical spin valve devices}

\author{Dietmar Czubak}
\email{czubak@pdi-berlin.de}
\author{Samuel Gaucher}
\author{Lars Oppermann}
\author{Jens Herfort}
\affiliation{Paul-Drude-Institut f\"ur Festk\"orperelektronik, Leibniz-Instiut im Forschungsverbund Berlin e. V., Hausvogteiplatz 5--7, 10117 Berlin, Germany}
\author{Klaus Zollner}
\author{Jaroslav Fabian}
\affiliation{Institute for Theoretical Physics, University of Regensburg, 93040 Regensburg, Germany}
\author{Holger T. Grahn}
\author{Manfred Ramsteiner}
\email{ramsteiner@pdi-berlin.de}
\affiliation{Paul-Drude-Institut f\"ur Festk\"orperelektronik, Leibniz-Instiut im Forschungsverbund Berlin e. V., Hausvogteiplatz 5--7, 10117 Berlin, Germany}

\begin{abstract}
We studied metastable $\alpha$-FeGe$_2$, a novel layered tetragonal material, embedded as a spacer layer in spin valve structures with ferromagnetic Fe$_3$Si and Co$_2$FeSi electrodes.
For both types of electrodes, spin valve operation is demonstrated with a metallic transport behavior of the $\alpha$-FeGe$_2$ spacer layer. The spin valve signals are found to increase both with temperature and spacer thickness, which is discussed in terms of a decreasing magnetic coupling strength between the ferromagnetic bottom and top electrodes. The temperature-dependent resistances of the spin valve structures exhibit characteristic features, which are explained by ferromagnetic phase transitions between 55 and 110~K. The metallic transport characteristics as well as the low-temperature ferromagnetism are found to be consistent with the results of first-principles calculations.
\end{abstract}

\maketitle

\section*{Introduction}

Vertical spin valves are essential building blocks for spintronic applications and valuable tools for fundamental research \citep{wolf2001spintronics,feng2017prospects} After the discovery of the giant magnetoresistance (GMR) effect \citep{dieny1991spin}, the exploration of the tunneling magnetoresistance (TMR) effect heralded the next era of spin valves \citep{moodera1995large}, peaking at a magnetoresistance ratio of 604\% at room temperature in a device structure with MgO as spacer material \citep{ikeda2008tunnel}. Nowadays, the spacer material between the ferromagnetic electrodes becomes more and more of interest in recent research activities aiming at multi-functionalities and tunabilities of spacer materials, rather then outperforming read-out efficiencies. Two-dimensional (2D) materials like transition metal dichalcogenides (TMDs) \citep{wang2015spin,iqbal2016room} or graphene \citep{singh2014negative} have become of major interest during the last years, because of their wide range of electronic characteristics including semiconductiing \citep{boker2001band} and superconducting \citep{zhang2016superconductivity} transport behavior as well as halfmetallic ferromagnetism \citep{tong2017half}. 

In this article, we introduce the novel layered material \mbox{$\alpha$-FeGe$_2$} as one of the few promising candidates regarding the search for two-dimensional spintronic materials. The successful synthesis of metastable \mbox{$\alpha$-FeGe$_2$} was only recently demonstrated utilizing a solid phase epitaxial process \citep{jenichen2018a}. First studies on the material by electron microscopy and synchrotron x-ray diffraction revealed a layered tetragonal structure (space group $P4mm$) which can be grown quasi-two-dimensional similar to MoS$_2$. Various physical properties and phenomena are proposed for \mbox{$\alpha$-FeGe$_2$} including magnetic phase transitions and high-$T_{\mathrm{C}}$ superconductivity \citep{stewart2011superconductivity, miiller2015protected}. For the counterpart \mbox{$\alpha$-FeSi$_2$}, which so far has been much more investigated, a wide tunability of the electronic and magnetic properties has been predicted, with nonmetallic transport and ferromagnetism being observed in strain-stabilized thin films \citep{cao2015a}. With a similar tunability of the physical properties, \mbox{$\alpha$-FeGe$_2$} films could be utilized both as ferromagnetic electrodes and barrier material for spintronic applications. Furthermore, the tuning of \mbox{$\alpha$-FeGe$_2$} might result in one of the rare 2D magnetic materials required for 2D spintronics \citep{gibertini2019a,gong2019a,cortie2019two}. However, the electronic and magnetic properties of \mbox{$\alpha$-FeGe$_2$} are so far basically unexplored from the experimental point of view. Since the synthesis of \mbox{$\alpha$-FeGe$_2$} includes the interdiffusion between amorphous Ge and an underlying \mbox{Fe$_3$Si} layer, investigations of the lateral transport and magnetometry measurements are impeded by the difficulty to avoid a remaining thin film of \mbox{Fe$_3$Si} underneath the \mbox{$\alpha$-FeGe$_2$} layer \citep{gaucher2017a,gaucher2018a,Terker2019:SST}.

Here, we utilized vertical transport in spin valve structures to shed light on the electronic and magnetic characteristics of embedded \mbox{$\alpha$-FeGe$_2$} as well as testing their potential for spintronic applications.

\section*{Experimental and Computational Details}

The investigated vertical spin valve devices are based on a trilayer structure, in which an \mbox{$\alpha$-FeGe$_2$} film serves as the spacer layer between a ferromagnetic bottom (FM1) and a ferromagnetic top (FM2) electrode. The trilayer structures were grown by a combination of low-temperature molecular beam epitaxy (MBE) and solid-phase epitaxy (SPE) on semi-insulating GaAs(001) substrates according to an approach described previously \citep{gaucher2017a,gaucher2018a}. The FM1 and FM2 films consist of either \mbox{Fe$_3$Si} or \mbox{Co$_2$FeSi}. The complete hybrid structures with \mbox{$\alpha$-FeGe$_2$} interlayers were found to be monocrystalline \citep{jenichen2018a,gaucher2017a,gaucher2018a}. An overview of the sequences and thicknesses of the individual films in the investigated trilayer structures is given in Table~\ref{samples}.

\begin{table}
\caption{\label{samples}Layer sequence of the trilayer structures used for the investigated spin valve devices and individual layer thicknesses.}
\begin{ruledtabular}
\begin{tabular}{ccccccc}
Device & FM1 & $d_1$ (nm) & interlayer & $t$ (nm) & FM2 & $d_2$ (nm)\\
\hline
\#1 & Fe$_3$Si & 36 & FeGe$_2$ & 4 & Fe$_3$Si & 12\\
\#2 & Fe$_3$Si & 36 & FeGe$_2$ & 6 & Fe$_3$Si & 12\\
\#3 & Fe$_3$Si & 36 & FeGe$_2$ & 8 & Fe$_3$Si & 12\\
\#4 & Fe$_3$Si & 36 & FeGe$_2$ & 6 & Co$_2$FeSi & 12\\
\end{tabular}
\end{ruledtabular}
\end{table}

To fabricate the spin valve devices, photolithography and wet etching were used to define square pillars with a surface area of 1~$\upmu$m$^2$ as shown in Figs.~\ref{device}(a) and \ref{device}(b). The pillars were contacted by a Ti/Au alloy on top of insulating \mbox{SiO$_2$}, both deposited by vapor deposition. For the magnetoresistance measurements, the device resistance was measured with a fixed current of \mbox{1~mA} using a three-terminal configuration as shown in Fig.~\ref{device}(a). The external magnetic field was applied along a \mbox{$\left\langle 110 \right\rangle$-direction} of the GaAs substrate, because the detected spin valve signals showed the highest amplitude along this direction.

The electronic structure calculations were performed by density functional theory (DFT)~\cite{Hohenberg1964:PRB}
with {\tt Quantum ESPRESSO}~\cite{Giannozzi2009:JPCM}.
Self-consistent calculations were carried out with the $k$-point sampling of $48\times 48\times 36$.
We performed open shell calculations that provide the spin-polarized ground state for bulk \mbox{$\alpha$-FeGe$_2$}. We used an energy cutoff for the charge density of $700$~Ry, and the kinetic energy cutoff for the wavefunctions was $80$~Ry for the fully relativistic pseudopotentials employing the projector augmented-wave method \cite{Kresse1999:PRB}, with the Perdew-Burke-Ernzerhof exchange correlation functional \cite{Perdew1996:PRL}. The atomic structure of bulk \mbox{$\alpha$-FeGe$_2$} was taken from recent experiments \cite{jenichen2018a, ASE}. More details about our DFT calculations are given in the Supplemental Material.

\section*{Results and Discussion}

\begin{figure}
\subfloat{
\includegraphics*[width=0.44\textwidth]{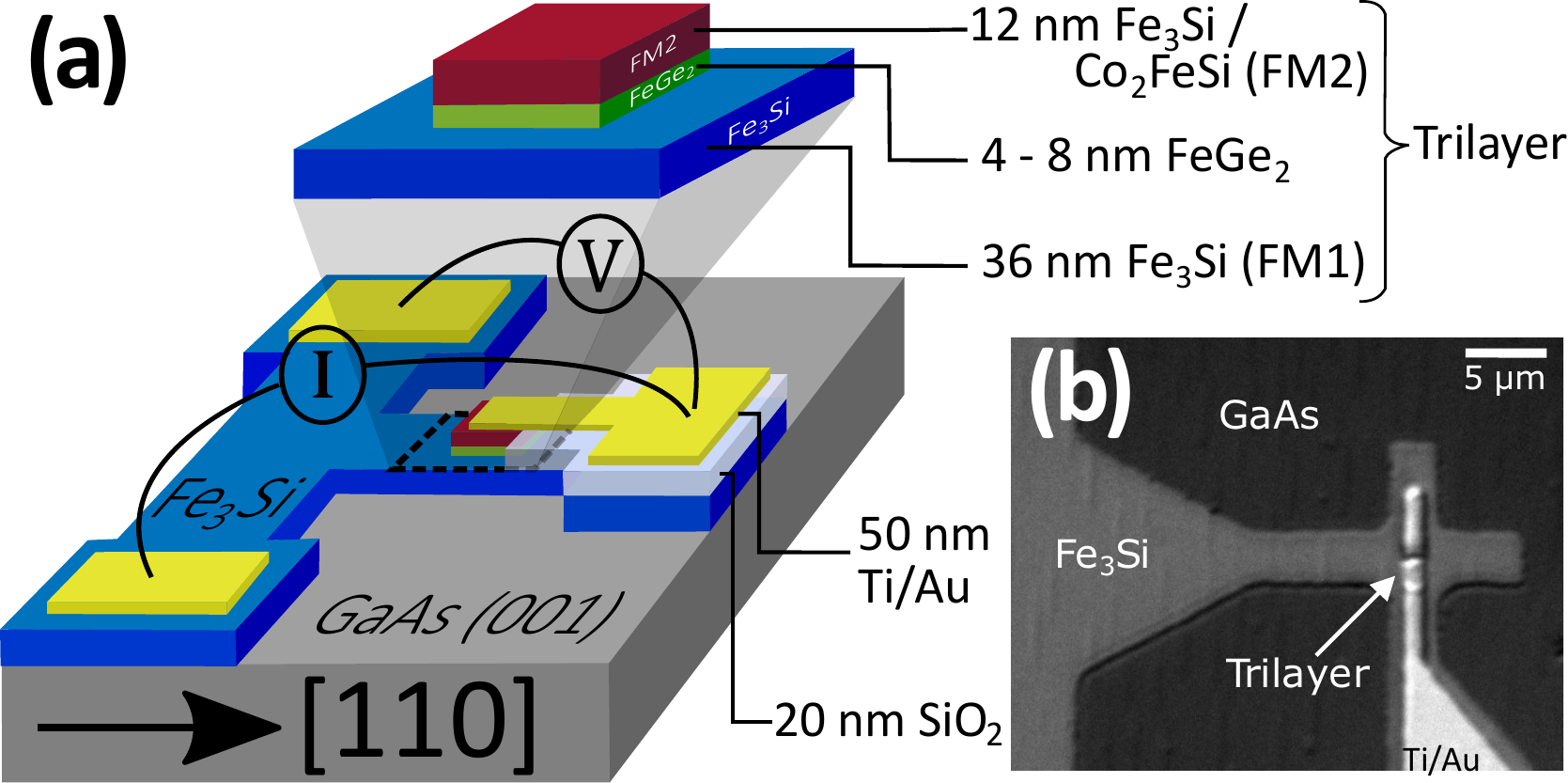}}
\vspace{\floatsep}
\subfloat{
\includegraphics[width=0.41\textwidth]{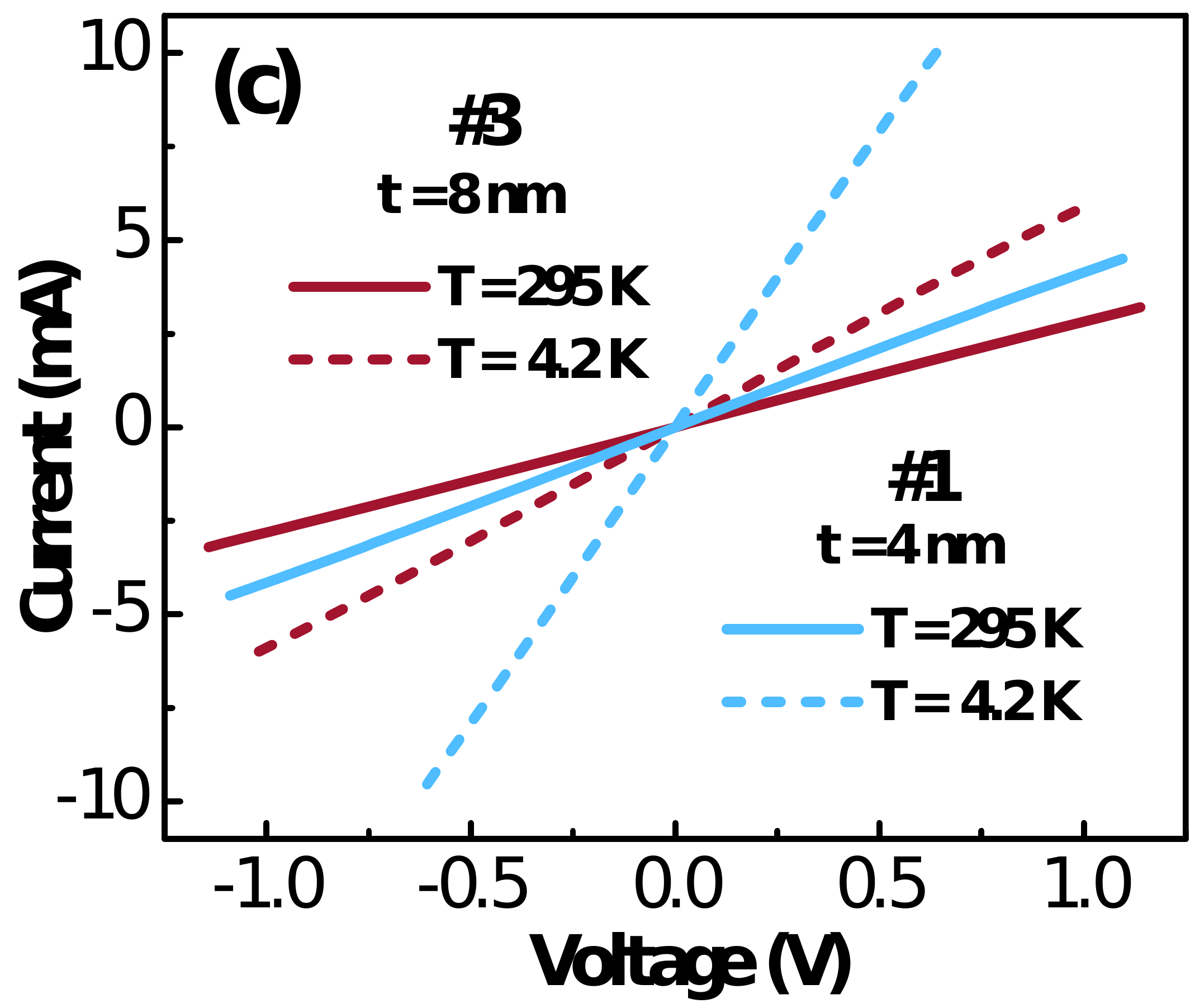}}
\caption{(a) Schematic diagram of the device structure and the configuration of the magnetoresistance measurements for vertical transport. (b) Optical micrograph of the spin valve device. (c) Current-voltage ($I$-$V$) characteristics for the devices \#1 and \#3 at temperatures of 295 (solid lines) and $4.2\,$K (dashed lines).}
\label{device}
\end{figure}

The current-voltage ($I$-$V$) characteristics for vertical transport through the devices \#1 and \#3 is shown in Fig.~\ref{device}(c). Similar results were obtained for the devices \#2 and \#4. Nearly perfect ohmic behavior was observed at room temperature and at low temperatures with no indication of tunneling or rectification, e.g., due to the formation of a Schottky barrier at the \mbox{$\alpha$-FeGe$_2$}/\mbox{FeSi$_3$} interfaces. The resistance area product ($RA$) exhibited the expected increase for increasing the \mbox{$\alpha$-FeGe$_2$} spacer thickness ($235\,\upOmega\, \upmu$m$^2$ for device \#1 and $352\,\upOmega\, \upmu$m$^2$ for device \#3) and decreased, when the temperature was reduced ($62\,\upOmega\, \upmu$m$^2$ and $163\,\upOmega\, \upmu$m$^2$ for device \#1 and \#3, respectively). Altogether, these findings clearly prove the metallic transport behavior of the \mbox{$\alpha$-FeGe$_2$} film in accordance with our DFT calculations (see below). Note that in the case of device resistances dominated by tunneling, much higher $RA$ values and no decrease of the resistance at low temperatures would be expected \citep{patino2015a}.

\begin{figure}
\subfloat{
\includegraphics*[width=0.41\textwidth]{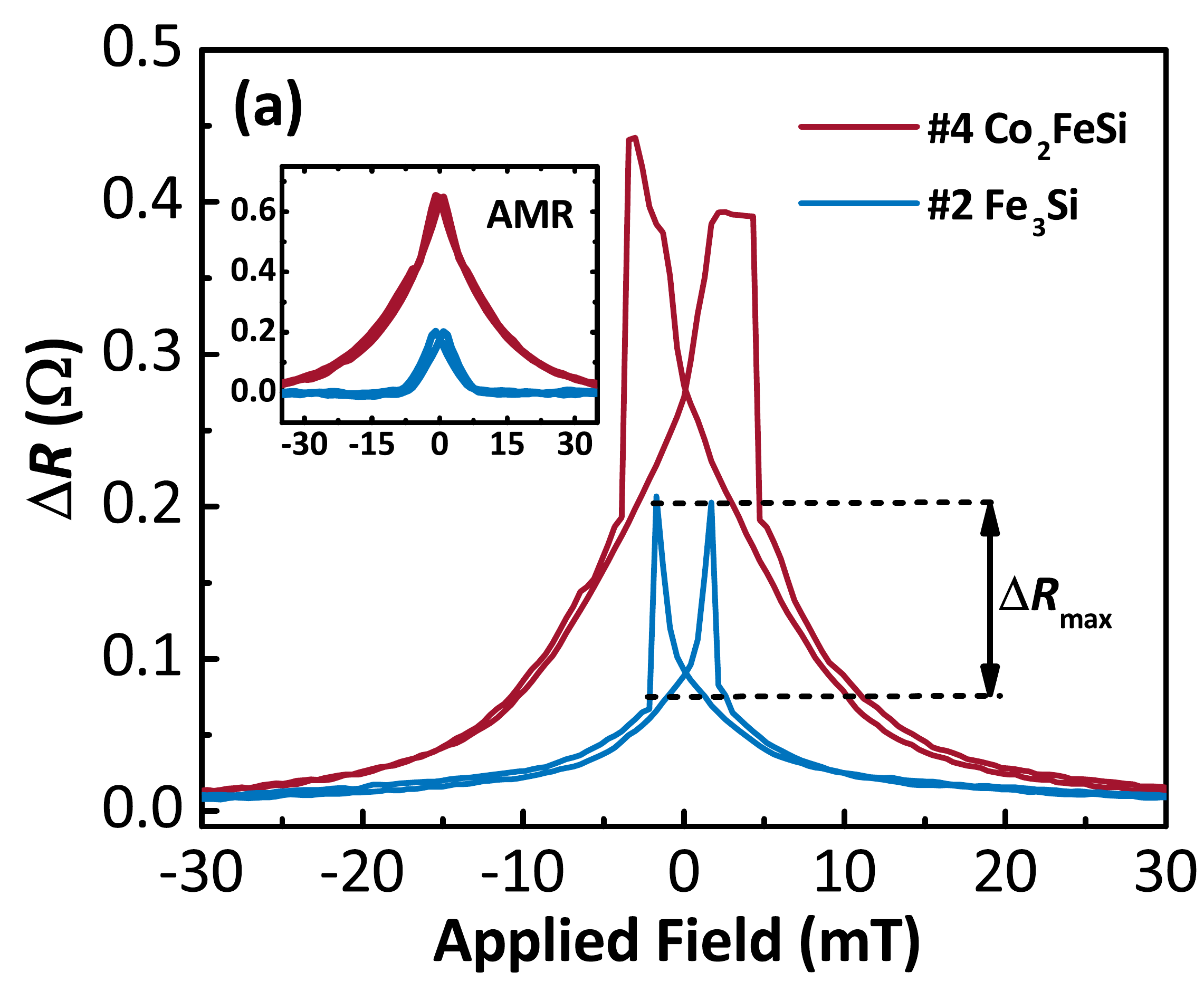}}
\\
\subfloat{
\includegraphics[width=0.41\textwidth]{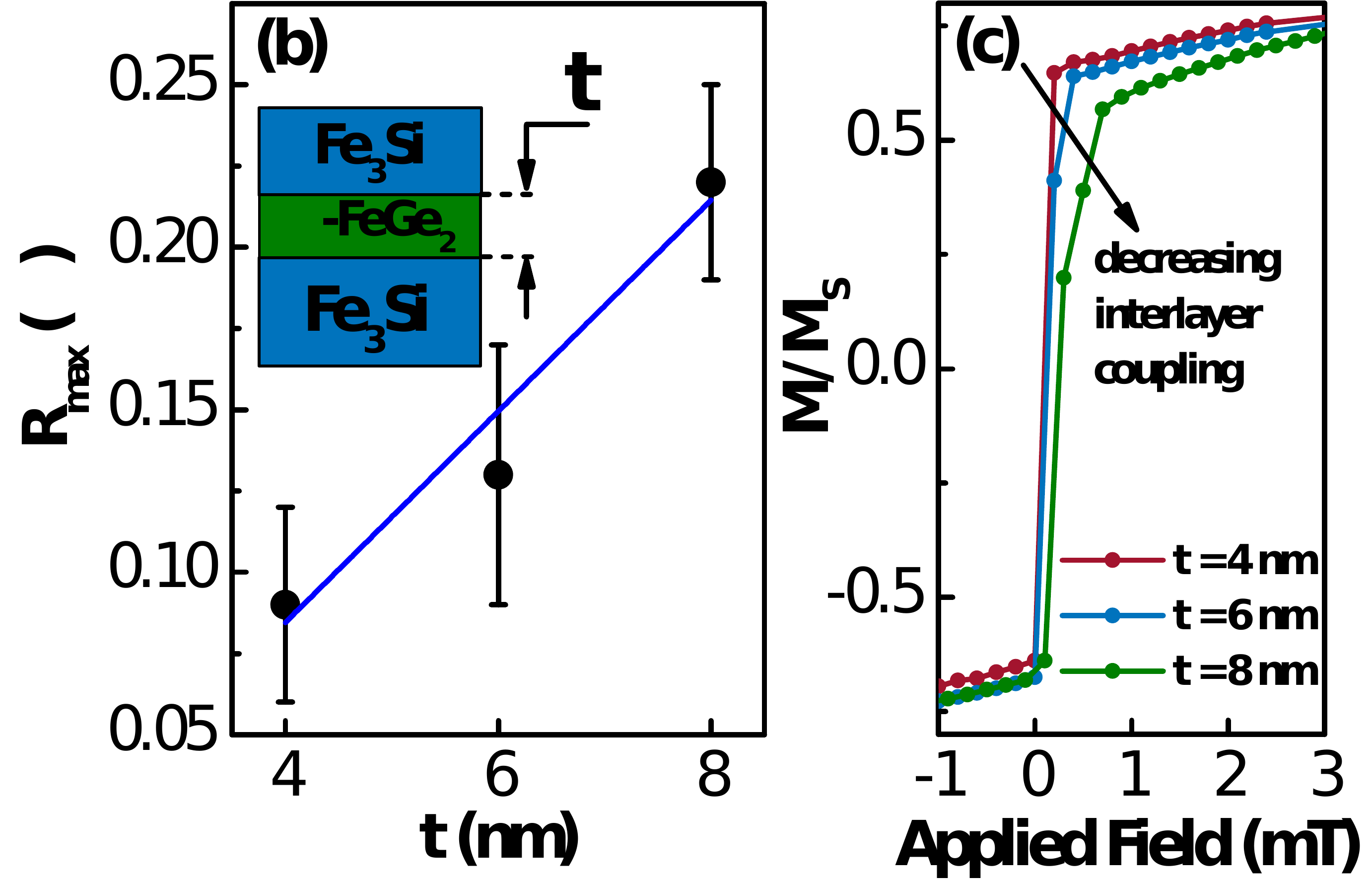}}
\caption{(a) Change in resistance ($\Delta R$) as a function of an external in-plane magnetic field (upward and downward sweeps) for the devices \#2 and \#4 along the $\left[ 110 \right]$-direction. The inset displays the typical anisotropic magnetoresistance (AMR) signals obtained for the lateral transport in the FeSi$_3$ stripes. (b) Spin valve signal $\Delta R_{\text{max}}$ as a function of the $\alpha$-FeGe$_2$ interlayer thickness $t$ for the devices \#1, \#2, and \#3. The solid line is a guide to the eye. (c) One-way normalized SQUID magnetization curves along the \mbox{$\left[ 110 \right]$-direction} (applied field swept form negative to positive fields) for Fe$_3$Si/\mbox{$\alpha$-FeGe$_2$}/Fe$_3$Si samples with different spacer thicknesses. (a-c) All the presented results were measured at room temperature.}
\label{valve}
\end{figure}

The spin transport through the trilayer structures in the different devices was studied
by examining the change in resistance $\Delta R(H) = R(H) - R_{\text{p}}$ during
upward and downward sweeps of an external magnetic field ($H$), where $R_{\text{p}}$
denotes the resistance in a large magnetic field. Figure~\ref{valve}(a) reveals
characteristic peaks in the change of the resistance ($\Delta R$) as signatures of
successful spin valve operation for the two different FM2 materials (\mbox{Co$_2$FeSi}
and \mbox{Fe$_3$Si}). The high- and low-resistance states correspond to the antiparallel
and parallel magnetization configurations (FM1 vs.$\,$FM2), respectively. The larger
widths of the spin valve signals observed for the devices with \mbox{Co$_2$FeSi} as top
electrode (device \#4) are due to the higher coercive field compared to
\mbox{Fe$_3$Si} \citep{herfort2003a,hashimoto2005a}.
Note that the spin valve signals are superimposed on the anisotropic
magnetoresistance (AMR) caused by the lateral transport in the \mbox{Fe$_3$Si} stripes
[cf. Fig.~\ref{device}(a)] \citep{bowen2005order, erekhinsky2013spin}. The occurrence of the AMR signal reflects the fact that \mbox{$\left\langle 100 \right\rangle$-directions} constitute the easy axes of magnetization in Fe$_3$Si whereas the external magnetic field is applied along a \mbox{$\left\langle 100 \right\rangle$-direction}. As a consequence the magnetization in the Fe$_3$Si stripes rotates by an angle of $45^\circ$ during a sweep from large to zero magnetic field.
For the determination of $\Delta R_{\text{max}}$, the AMR contribution has been taken
into account as a background signal. The larger spin valve signal detected for device
\#4 ($\Delta R_{\text{max}} = $~0.26~$\Omega$) compared to that of device \#2
($\Delta R_{\text{max}} = $~0.13~$\Omega$) is attributed to the higher spin
polarization in \mbox{Co$_2$FeSi} \citep{ramsteiner2008a,hamaya2012a}. The corresponding
relative magnetoresistances $\Delta$MR$ = \Delta R_{\text{max}}/R_{\text{p}}$ are
$0.10\%$ and $0.17\%$ for devices \#2 and \#4, respectively. Note, however, that these
values are influenced by the background resistance $R_{\text{FM1}}$ originating from
lateral transport in the \mbox{Fe$_3$Si} bottom layer which contributes to the magnitude of $R_{\text{p}}$.

The spin valve signal $\Delta R_{\text{max}}$ has been found to become progressively larger with increasing the thickness of the \mbox{$\alpha$-FeGe$_2$} interlayer as shown in Fig.~\ref{valve}(b). Such a monotonic increase has been previously observed for metallic spacer layers and attributed to a thickness dependent magnetic coupling between the electrodes FM1 and FM2 \citep{rijks1994a,kools1995a,leal1996a}. Magnetic interlayer coupling between ferromagnetic electrodes has been discussed in the literature in terms of a strong magnetostatic interaction, the density of pinholes, and Néel's orange-peel coupling \citep{dieny1991a,kools1995b}. For comparatively thin spacer layers, the interlayer coupling is expected to be relatively strong and to favor a parallel alignment of the ferromagnetic electrodes as well as a simultaneous magnetization reversal. Consequently, a complete antiparallel alignment during magnetic field sweeps is prevented which results in reduced spin valve signals. With increasing spacer thickness, the interlayer coupling strength decreases and thus also its detrimental influence on the spin valve signal, in accordance with our experimental observation. Our explanation is supported by the magnetometry measurements shown in
Fig.~\ref{valve}(c). The magnetization reversals measured by a superconducting
quantum interference device (SQUID) exhibit a clear dependence on the
spacer thickness in Fe$_3$Si/\mbox{$\alpha$-FeGe$_2$}/Fe$_3$Si trilayer structures. The kink which develops between $M$/$M_{\mathrm{s}} = 0$ and $M$/$M_{\mathrm{s}} = 0.5$ with increasing
spacer thickness indicates a progressing magnetic decoupling of the
ferromagnetic layers \citep{schanzer2005magnetic, liu2009temperature, liu2010strong}. For even larger spacer thicknesses, this
kink is expected to develop into a step in the magnetization curve as a
signature of fully independent magnetization reversals in the two
decoupled ferromagnetic electrodes. Note that the observed coercive
fields in the SQUID magnetization curves are much smaller than the range
of switching fields at which the spin valve signals occur (see Fig.~\ref{valve}(a)). This discrepancy is most likely due to the shape anisotropy
induced during the microstructuring of the devices and additional
demagnetization field effects from impurities at the contact edges
\citep{fluitman1973influence, kryder1980magnetic, lee1999effect}. As a consequence, the range of spacer thicknesses at which the
transition from strong to weak magnetic interlayer coupling occurs is
also expected to be somewhat different for large-area SQUID samples compared to microstructured spin valve devices.
Although, the observed spacer-thickness dependence of $\Delta R_{\mathrm{max}}$ is consistent with metallic transport behavior, we cannot fully rule out additional spin filtering phenomena in the case of possible tunneling processes \citep{worledge2000magnetoresistive}. Beyond the regime of strong interlayer coupling, a decrease of the spin valve signal with increasing spacer thickness is expected according to a finite spin diffusion length in the spacer material \citep{speriosu1993a,kools1995a}. Consequently, our result demonstrates that the magnetic interlayer coupling dominates over the influence of spin relaxation in the spacer for the entire range of investigated thicknesses, indicating extraordinarily strong magnetic coupling effects and a large spin diffusion length in \mbox{$\alpha$-FeGe$_2$}. However, the investigated spacer thicknesses are rather large compared to the previously studied cases of magnetic interlayer coupling between ferromagnetic electrodes \citep{dieny1991a,rijks1994a,kools1995a,kools1995b,leal1996a}. Therefore, interfacial exchange coupling as an additional influence on the relative alignment of the magnetization in the ferromagnetic electrodes has to be taken into account \citep{pankratova2015model}. For the occurrence of this additional mechanism, antiferromagnetic order in the \mbox{$\alpha$-FeGe$_2$} spacer has to be assumed which would be in accordance with energetic considerations from DFT calculations (see supplemental) and the magnetic characteristic of the stable phase \mbox{$\beta$-FeGe$_2$} \citep{mason1997itinerant}. However, further work is necessary to clarify this point.

\begin{figure}
\includegraphics*[width=0.45\textwidth]{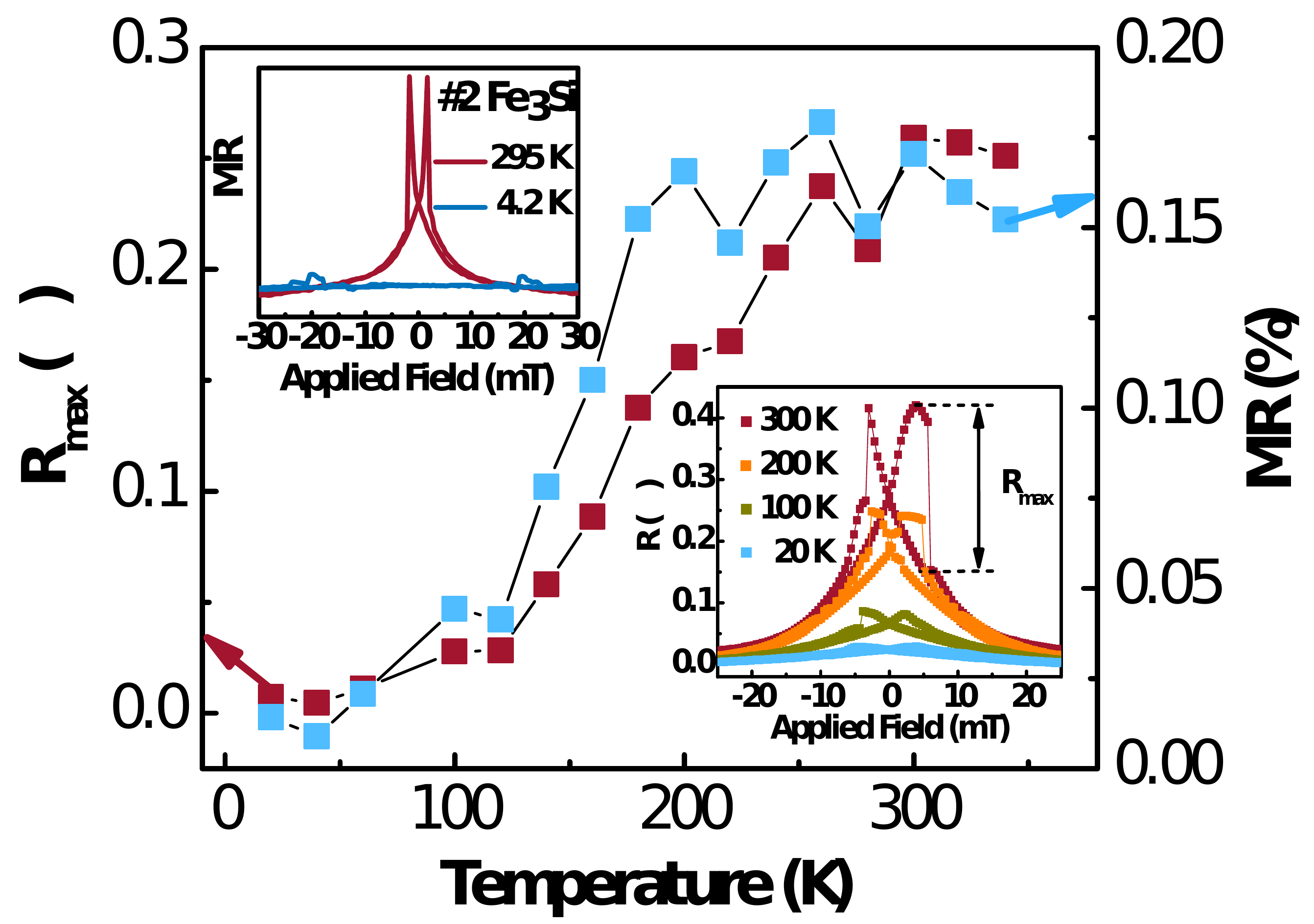}
\caption{Temperature dependence of the spin valve signal $\Delta R_{\text{max}}$ and the relative magnetoresistance MR for device \#4.
Bottom right inset: Spin valve signals of device \#4 at different temperatures. Top left inset: Spin valve signals (magnetoresistances) of device \#2 at 295~K (red) and 4.2~K (blue).}
\label{sv_temp}
\end{figure}

In striking contrast to the commonly observed behavior for vertical spin valves \citep{kimura2012a,iqbal2016room,tsymbal2003a}, the spin valve signal of our devices vanishes with decreasing temperature as shown in Fig.~\ref{sv_temp}. In fact, no characteristic spin valve signal could be detected for temperatures below 100~K. The initial decrease between room temperature and 100~K can be explained by an increasing magnetic interlayer coupling strength along the lines of the discussion above on the spacer thickness dependence. Indeed, examples of an increasing magnetic coupling strength with decreasing temperature have been reported previously \citep{hu2019a,persat1997a,zhang1994a}. An increase in the interlayer coupling strength is also indicated by the slight decrease of the coercive fields at which the spin valve signals occur when lowering the temperatures (see bottom right inset of Fig.~\ref{sv_temp}). The complete disappearance of the spin valve signal below 100~K is attributed to a ferromagnetic phase transition in the spacer material \mbox{$\alpha$-FeGe$_2$}. The same quenching of the spin valve signal below 100~K has been observed for devices with Fe$_3$Si top electrodes (see top left inset of Fig.~\ref{sv_temp} for the case of device \#2). Our DFT calculations determined the ferromagnetic phase to be the energetically most favorable one at low temperature (see discussion below). In the case of a fully ferromagnetic trilayer device structure below 100~K, a particularly strong interlayer coupling is expected leading to a simultaneous magnetization reversal in the electrodes FM1 and FM2 as well as the \mbox{$\alpha$-FeGe$_2$} spacer which excludes the occurrence of a spin valve signal. Furthermore, the small spin diffusion length commonly observed in ferromagnetic materials \citep{bass2007spin} might also contribute to the quenching of spin valve signal below the Curie temperature of \mbox{$\alpha$-FeGe$_2$}. Note that magnetic phase transitions below room temperature were already reported for the thermodynamically stable \mbox{$\beta$-FeGe$_2$} \citep{corliss1985magnetic,men2019magnetic} as well as for \mbox{FeGe$_2$} nanowires \citep{tang2017dimensionality}. Furthermore, successful spin valve operation has been achieved also for the trilayer system Fe$_3$Si/FeSi$_2$/Fe$_3$Si with a similar decrease of the device signal below 80~K \citep{asai2014fabrication,ishibashi2016temperature}.

\begin{figure}
\subfloat{
\includegraphics*[width=0.40\textwidth]{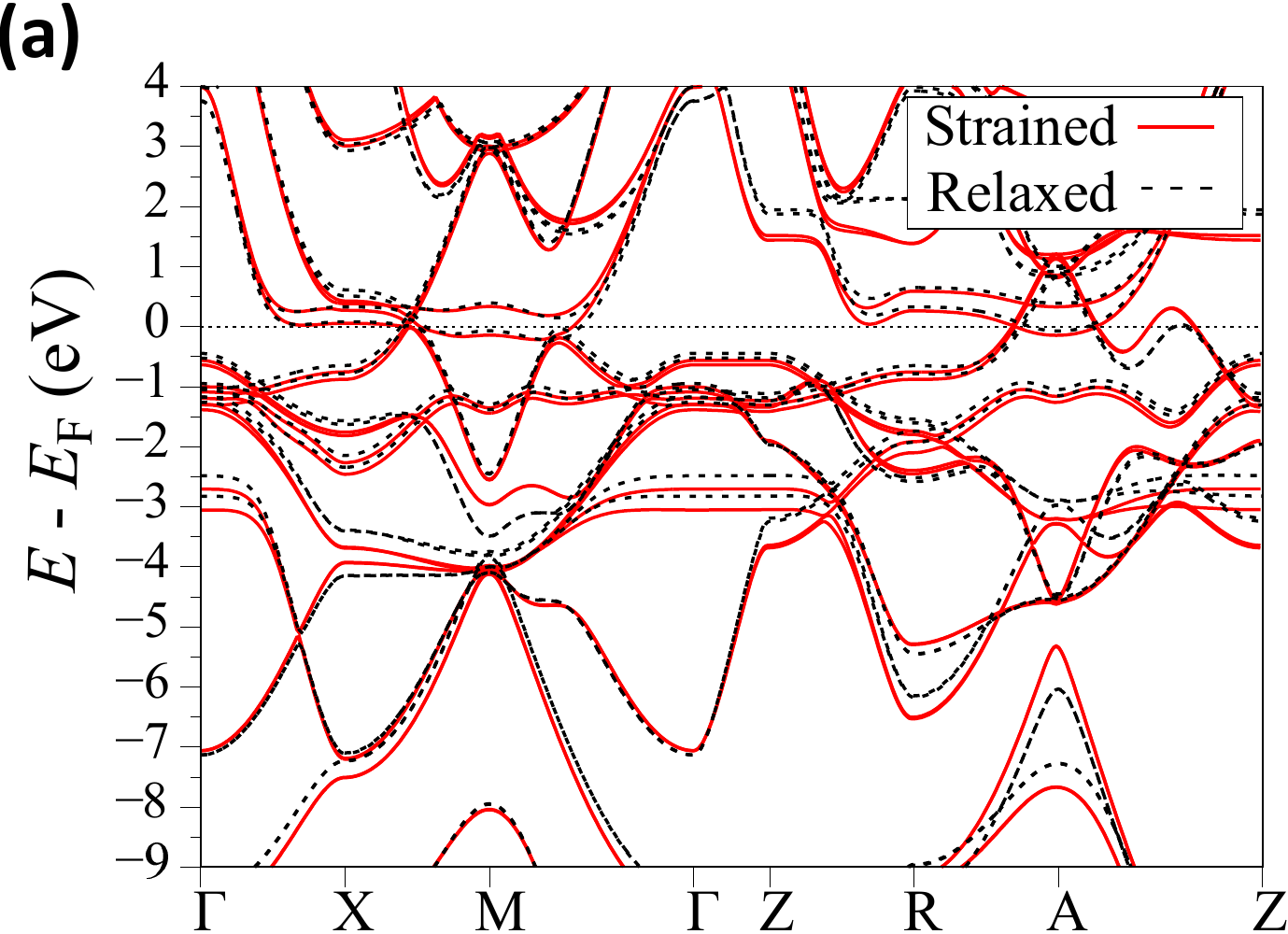}}
\\
\subfloat{
\includegraphics[width=0.41\textwidth]{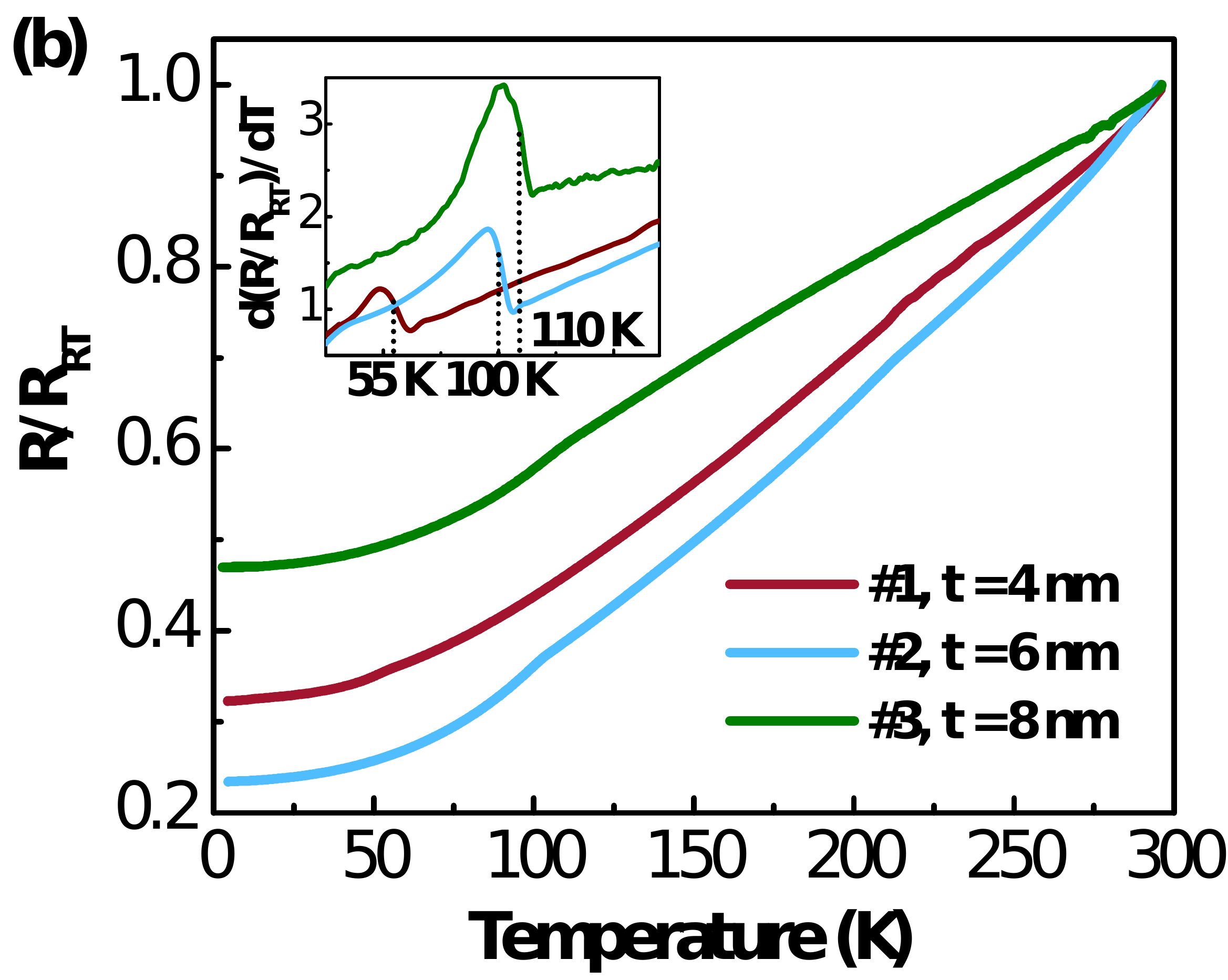}}
\caption{(a) Calculated band structure of bulk $\alpha$-FeGe$_2$ for
the experimentally determined strained lattice structure according to
Ref.~\onlinecite{jenichen2018a} (solid line) and for the fully relaxed lattice structure determined by DFT (dashed lines). (b) Resistance normalized to its room temperature value ($R/R_\text{RT}$) for  devices \#1, \#2, and \#3 as a function of temperature. The inset displays the corresponding temperature derivatives of the normalized resistances ($\text{d}(R/R_{\text{RT}})/(\text{d}T)$).}
\label{res_temp}
\end{figure}

The electronic band structure of isolated bulk \mbox{$\alpha$-FeGe$_2$} resulting from DFT calculations is shown in Fig.~\ref{res_temp}(a). Since the properties of
thin epitaxial films are often influenced by strain, the calculations
were performed for different lattice parameters: the fully relaxed structure
obtained by DFT and the experimentally determined strained geometry (see
Supplemental Material). The resulting overall band structure features
are nearly the same for both configurations and agree reasonably well
with the one reported in Ref.~\onlinecite{jenichen2018a}. Additional calculations carried out for the
conditions of uniaxial strain in the $\left<001\right>$- and $\left<110\right>$-directions up to $1\%$
do not lead to significant changes of these band structure features. The
absence of a band gap at the Fermi level $E_{\text{F}}$ confirms our findings
regarding the metallic transport characteristics (see above).

Regarding the magnetic ground state of \mbox{$\alpha$-FeGe$_2$}, we considered three
different configurations: (i) paramagnetic (PM) phase, where the atoms
are considered to have no collective magnetization axis; (ii) ferromagnetic (FM) phase, in
which all Fe atoms have aligned magnetic moments along the c-axis; and (iii)
antiferromagnetic (AFM) phase, where the ferromagnetic Fe sheets within \mbox{$\alpha$-FeGe$_2$} form sublattices with antiparallel order. From the total energy analysis of these
cases, we found that the FM phase is the energetically most favorable
ground state. The PM phase was about 56~meV higher in energy (or about 7~meV per unit cell) compared to the FM phase, while the AFM phase was only
about 2.9~meV higher in energy (or about 0.36~meV per unit cell). For
the DFT-relaxed structure, the FM phase was about 35~meV (19~meV) lower
in energy than the PM (AFM) phase. Again, we found that the FM phase
should be the ground state. These results indicate the possibility to
tune the magnetic ground state of \mbox{$\alpha$-FeGe$_2$} via the strain state. Note
that the strain in \mbox{$\alpha$-FeGe$_2$} might also depend on temperature because
of different thermal expansion coefficients of the film and the GaAs
substrate \citep{thomas2003interplay, mohanty2003effect}. This fact together with the small energy separations
between the ground states of the three configurations suggest a
relatively high probability for low-temperature magnetic phase
transitions.

Indeed, the temperature dependence of the device resistances exhibit characteristic features which can be attributed to low-temperature ferromagnetic phase transitions. Figure~\ref{res_temp}(b) displays the resistance normalized by the resistance at room temperature $R_{\text{RT}}$ as a function of temperature for the devices \#1, \#2, and \#3 in the magnetic virgin state with no cooling field applied. Characteristic changes in the $R/R_{\text{RT}}(T)$ curvature occur at distinct temperatures for the individual devices. The corresponding peaks in the temperature derivatives [cf. inset in Fig.~\ref{res_temp}(b)] can be explained by spin disorder scattering which starts to decrease below the Curie temperature $T_{\text{C}}$, where a phase change into a ferromagnetic state occurs. The accordingly determined Curie temperatures reveal thickness-dependent phase transitions occurring at 55, 100, and 110~K for spacer thicknesses of 4, 6, and 8~nm, respectively. Note, that the phase transition could not be identified for device \#4, most likely due to the comparatively large series resistance of the Co$_2$FeSi top electrode \citep{herfort2004structural, hashimoto2005thermal}. This result provides strong support for our explanation regarding the absence of a spin valve signal at temperatures below 100~K observed for device \#4 (cf. Fig.~\ref{sv_temp}). The determination of the ferromagnetic phase transition temperature for \mbox{$\alpha$-FeGe$_2$} spacer layers embedded in trilayer structures with ferromagnetic electrodes constitutes an achievement which is difficult to obtain by other means.

\section*{Summary and Conclusions}

The layered tetragonal material \mbox{$\alpha$-FeGe$_2$} embedded as a spacer layer in vertical spin valve structures exhibits metallic transport behavior. Successful spin valve operation is demonstrated for structures with ferromagnetic Fe$_3$Si (bottom and top) and \mbox{Co$_2$FeSi} (top) electrodes. An enhancement of the spin valve signals with increasing temperature and spacer layer thickness is attributed to a decreasing magnetic interlayer coupling between the ferromagnetic bottom and top electrodes. Characteristic features in the temperature derivatives of the device resistances are assigned to a ferromagnetic phase transition between 55 and 110~K. Both the metallic characteristics and the ferromagnetic ground state at low temperatures are confirmed by DFT calculations.

\section*{Acknowledgments}

K. Z. and J. F. were supported by DFG SFB 1277 (314695032). We gratefully acknowledge the technical support by Walid Anders and Angela Riedel as well as the critical reading of the manuscript by Alberto Hern\'andez-M\'inguez.


%

\end{document}


\title{Supplemental Material: Electronic and magnetic properties of $\alpha$-FeGe$_2$ films embedded in vertical spin-valve devices}

\author{Dietmar Czubak}
\email{czubak@pdi-berlin.de}
\author{Samuel Gaucher}
\author{Jens Herfort}
\affiliation{Paul-Drude-Institut f\"ur Festk\"orperelektronik, Leibniz--Instiut im Forschungsverbund Berlin e. V., Hausvogteiplatz 5--7, 10117 Berlin, Germany}
\author{Klaus Zollner}
\author{Jaroslav Fabian}
\affiliation{Institute for Theoretical Physics, University of Regensburg, 93040 Regensburg, Germany}
\author{Holger T. Grahn}
\author{Manfred Ramsteiner}
\email{ramsteiner@pdi-berlin.de}
\affiliation{Paul-Drude-Institut f\"ur Festk\"orperelektronik, Leibniz--Instiut im Forschungsverbund Berlin e. V., Hausvogteiplatz 5--7, 10117 Berlin, Germany}

	\begin{abstract}
		{}
	\end{abstract}
	
	\maketitle
	\onecolumngrid
	
\section{First-principles calculations}
\subsection{Computational Details}

The calculations of the electronic structure and the structural relaxation of
our geometries are performed with density functional theory (DFT)~\cite{Hohenberg1964:PRB}
using \textsc{Quantum Espresso} \cite{Giannozzi2009:JPCM}.
Self-consistent calculations are performed with the $k$-point
sampling of $48\times 48\times 36$ for bulk FeGe$_2$.
We use an energy cutoff for the charge density of $700$~Ry, and
the kinetic energy cutoff for the wavefunctions is $80$~Ry for the scalar relativistic pseudopotential
with the projector augmented-wave method \cite{Kresse1999:PRB} based on the
Perdew-Burke-Ernzerhof exchange correlation functional \cite{Perdew1996:PRL}.
When spin-orbit coupling (SOC) is included, the fully relativistic versions of the pseudopotentials are used.
Structural relaxations are calculated with a quasi-Newton algorithm based on the
trust radius procedure, until all components of all forces
are reduced below $10^{-4}$~[Ry/$a_0$], where $a_0$ denotes the Bohr radius.
    \begin{figure}[b]
    	\includegraphics[width=0.85\textwidth]{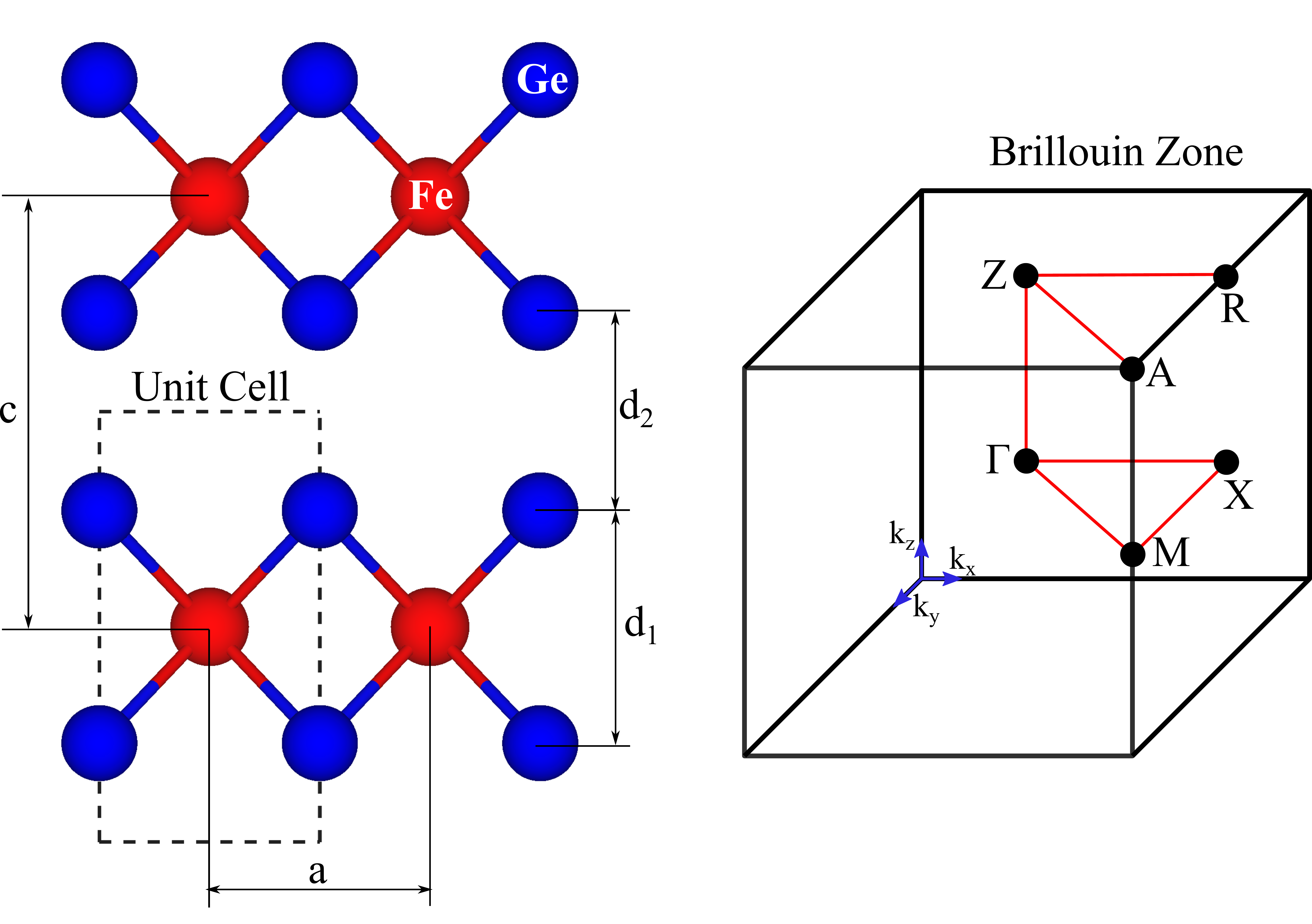}
    	\caption{Left: Side view of the geometry of the FeGe$_2$ crystal.
    	Red and blue spheres correspond to Fe and Ge atoms, respectively. The lattice structure is determined by
    	the lattice constants $a$ and $c$ as well as distances $d_1$ and $d_2$. One unit cell is emphasized by the dashed rectangle.
    	Right: First Brillouin zone of the FeGe$_2$ crystal including the high-symmetry points $\Gamma$, X, M, A, R, and Z.}
    	\label{fig:geometry}
    \end{figure}

\subsection{Geometry}

The geometry of FeGe$_2$ is shown on the left side of Fig. \ref{fig:geometry}, which is set up according to Jenichen \emph{et al.} \cite{Jenichen2018:PRM} with the help of the Atomic Simulation Environment (ASE) \cite{ASE}.
FeGe$_2$ crystallizes in the space group $P4mm$ (No. $99$) with lattice constants $a = 2.827~\text{\AA}$ as
$c = 5.517~\text{\AA}$ as well as distances $d_1 = 2.840~\text{\AA}$ and $d_2 = 2.677~\text{\AA}$.
So far, FeGe$_2$ has only been grown via solid-phase epitaxy in an Fe$_3$Si or Co$_2$FeSi environment \cite{Jenichen2018:PRM, Terker2019:SST, Gaucher2018:SST}.
Thus, the experimentally determined lattice constants are strained ones.
Unfortunately, this quasi-layered phase of FeGe$_2$ can, at the moment, only be grown in this special experimental setup.
For this experimentally determined structure, the calculated magnetic moments are $0.1777~\mu_{\text{B}}$ for Fe and $-0.0067~\mu_{\text{B}}$ for Ge.

If we fully relax the lattice structure and internal coordinates of bulk FeGe$_2$ derived from the calculated energy minimum of the lattice,
we obtain the lattice constants $a = 2.8525~\text{\AA}$ and
$c = 5.4118~\text{\AA}$ and distances $d_1 = 2.9191~\text{\AA}$ and $d_2 = 2.4927~\text{\AA}$.
In the calculated relaxed structure, the calculated magnetic moments are $0.1823 \mu_{\text{B}}$ for Fe and $-0.0066 \mu_{\text{B}}$ for Ge.

\subsection{Band Structure Results}

\begin{figure}[b]
    	\includegraphics[width=0.93\textwidth]{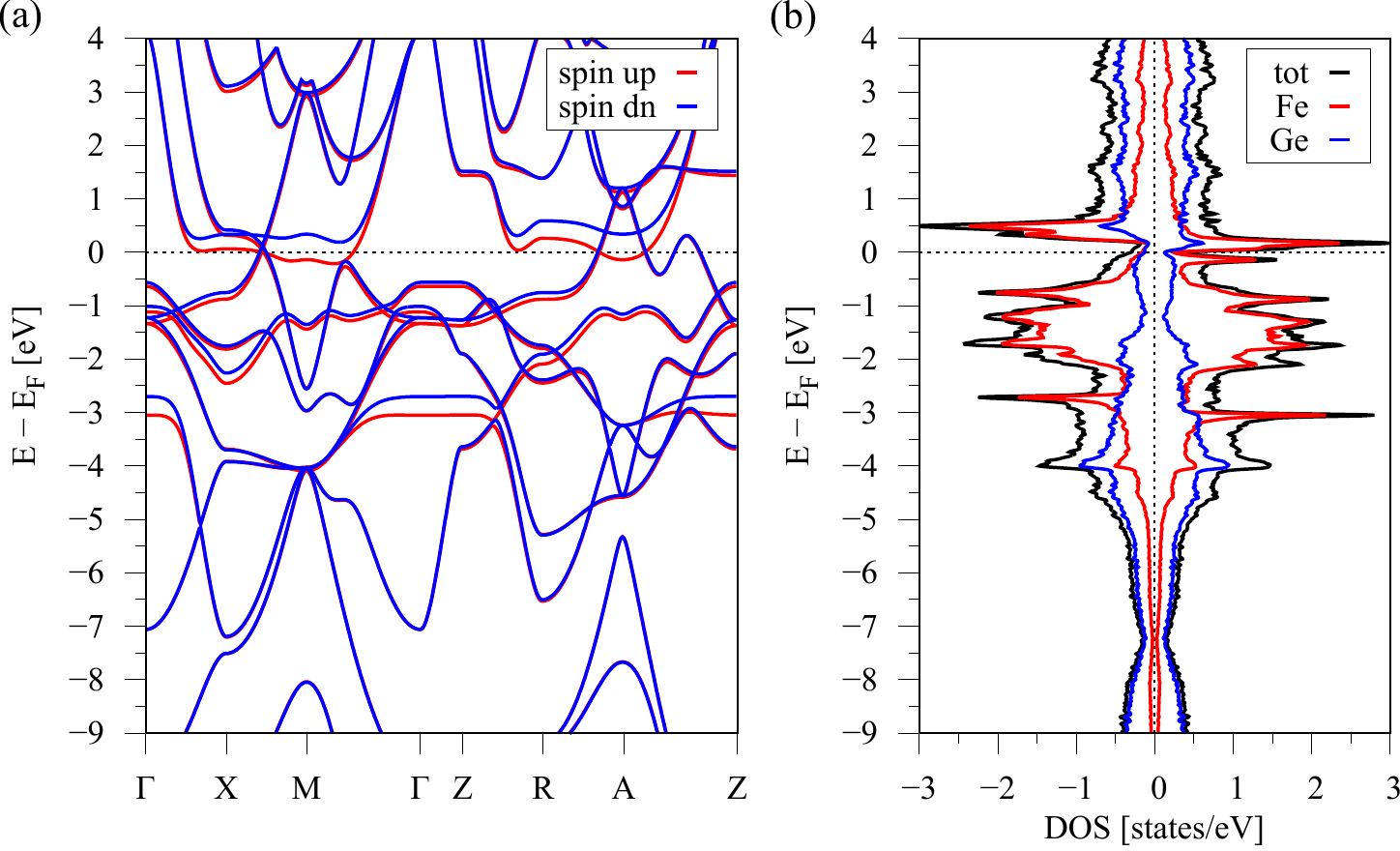}
    	\caption{(a) Calculated band structure using DFT of FeGe$_2$ without SOC. Bands marked in red (blue) correspond to spin-up (spin-down) states.
    	(b) Corresponding spin- and atom-resolved DOS. }
    	\label{fig:bands_noSOC}
    \end{figure}

\begin{figure}[t]
    	\includegraphics[width=0.74\textwidth]{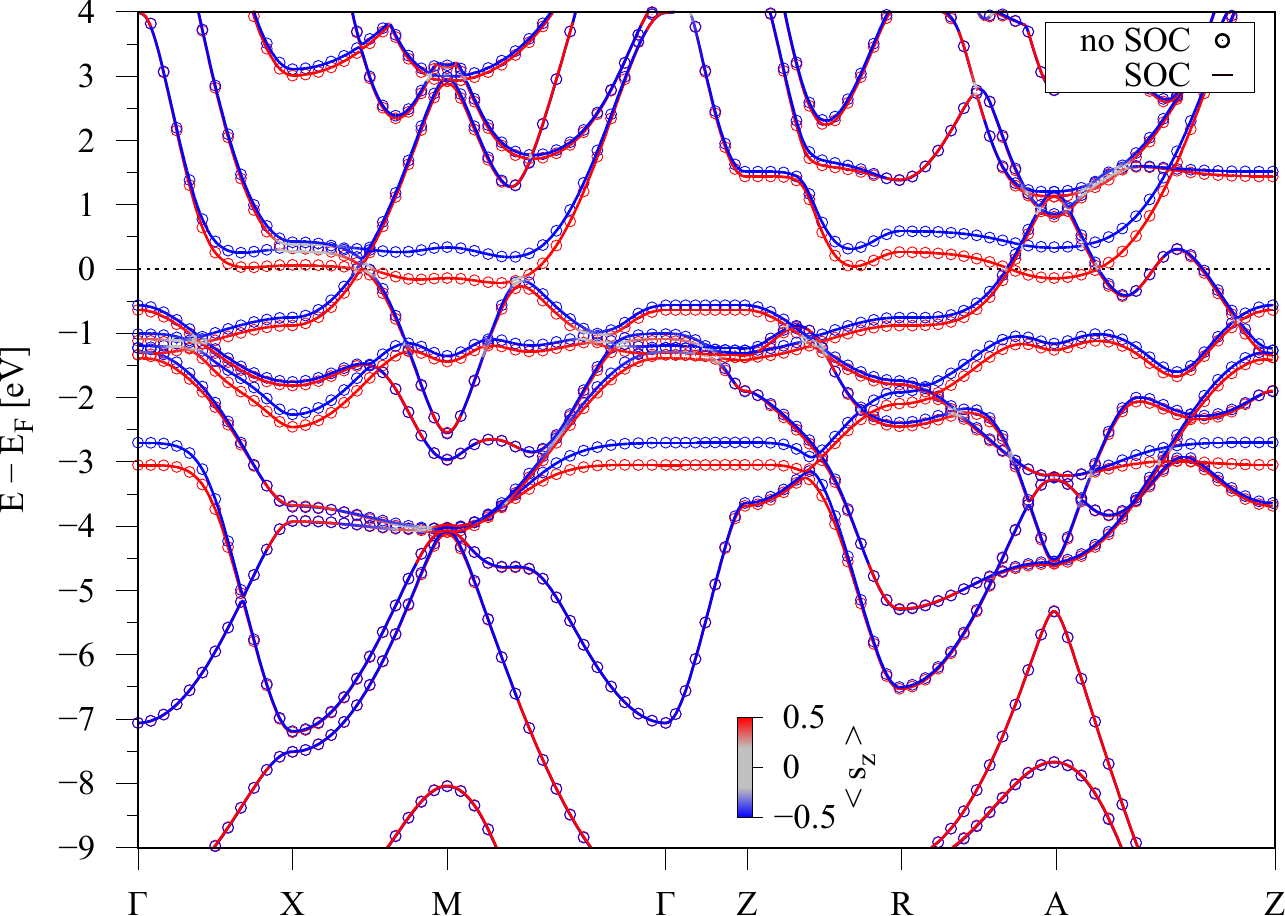}
    	\caption{Calculated band structure using DFT of FeGe$_2$ without (symbols) and with (solid lines) SOC for a spin quantization
    	axis along [001]-direction. The color scale indicates the $s_z$ spin expectation value.  }
    	\label{fig:bands_soc}
    \end{figure}

In the following, we discuss results obtained for the experimentally determined lattice structure of FeGe$_2$,
if not indicated otherwise.
In Fig. \ref{fig:bands_noSOC}, we show the calculated band structure as well as the spin- and atom-resolved
density of states (DOS) of the FeGe$_2$ crystal without SOC.
The overall band structure looks similar to the one of $\alpha$-FeSi$_2$ \cite{Cao2015:PRL},
which has roughly the same crystal structure as FeGe$_2$. In Fig. \ref{fig:bands_soc}, the calculated band structures of FeGe$_2$ including SOC is compared with the one without SOC.
The results do essentially not differ from each other so that SOC has a negligible effect on the band structure.

    \begin{figure}[b]
    	\includegraphics[width=0.7\textwidth]{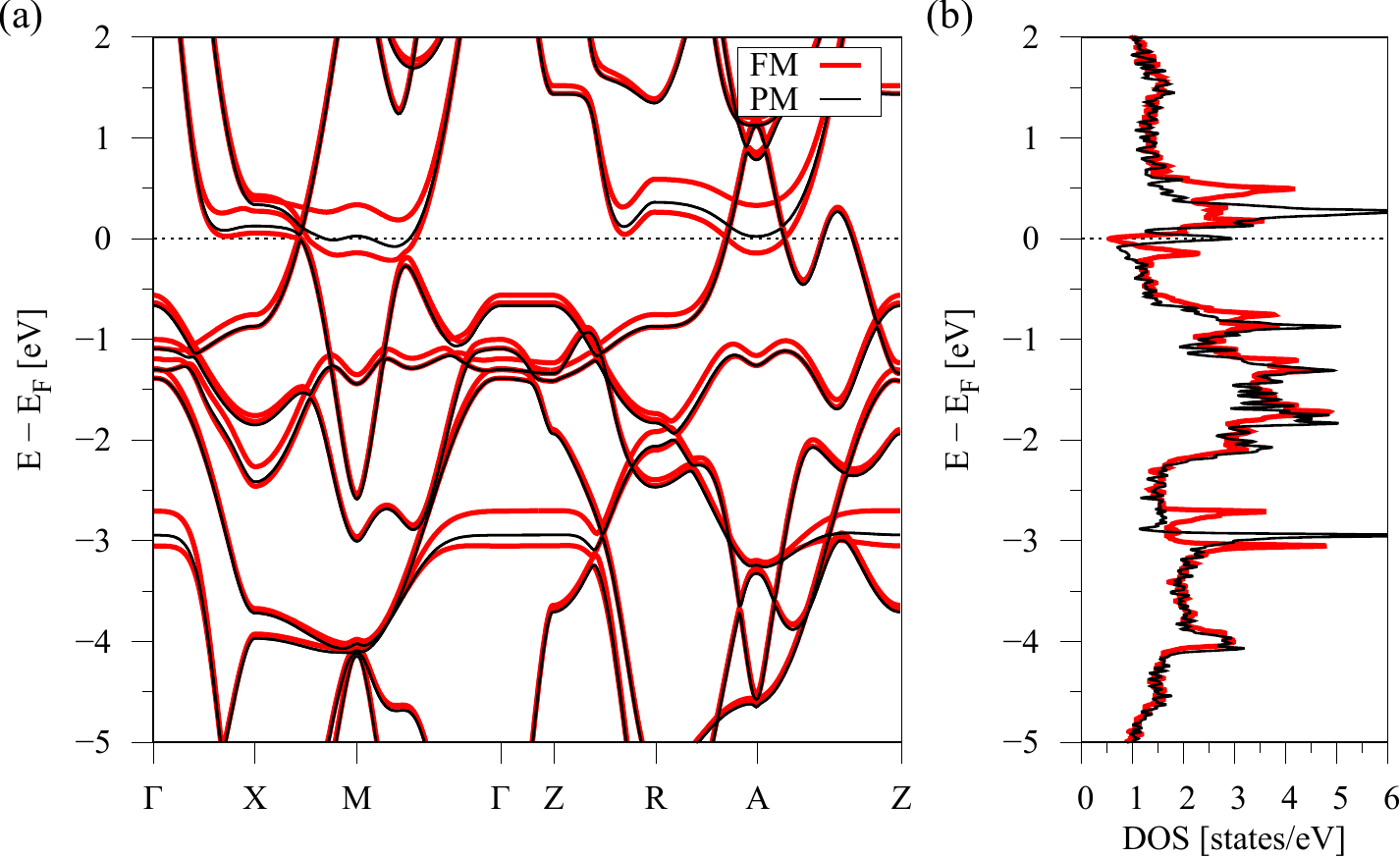}
    	\caption{(a) Calculated band structure and (b) corresponding total DOS using DFT of
    	FeGe$_2$ with SOC in the ferromagnetic (red lines)
    	and in the paramagnetic (black lines) state. }
    	\label{fig:bands_compare_FM_and_PM}
    \end{figure}
In Fig. \ref{fig:bands_compare_FM_and_PM}, the calculated band structure and total DOS are shown for FeGe$_2$ in the
ferromagnetic and paramagnetic phase. Especially in the total DOS, there is a characteristic difference at the Fermi level
between both phases. For the ferromagnetic phase, the total DOS shows a dip, while for the paramagnetic one there is a peak,
which again is similar to one for $\alpha$-FeSi$_2$ \cite{Cao2015:PRL}.
Comparing the total energies, we find that the ferromagnetic state of FeGe$_2$ is about 7~meV lower in energy than the
paramagnetic state.

        \begin{figure}[t]
    	\includegraphics[width=0.75\textwidth]{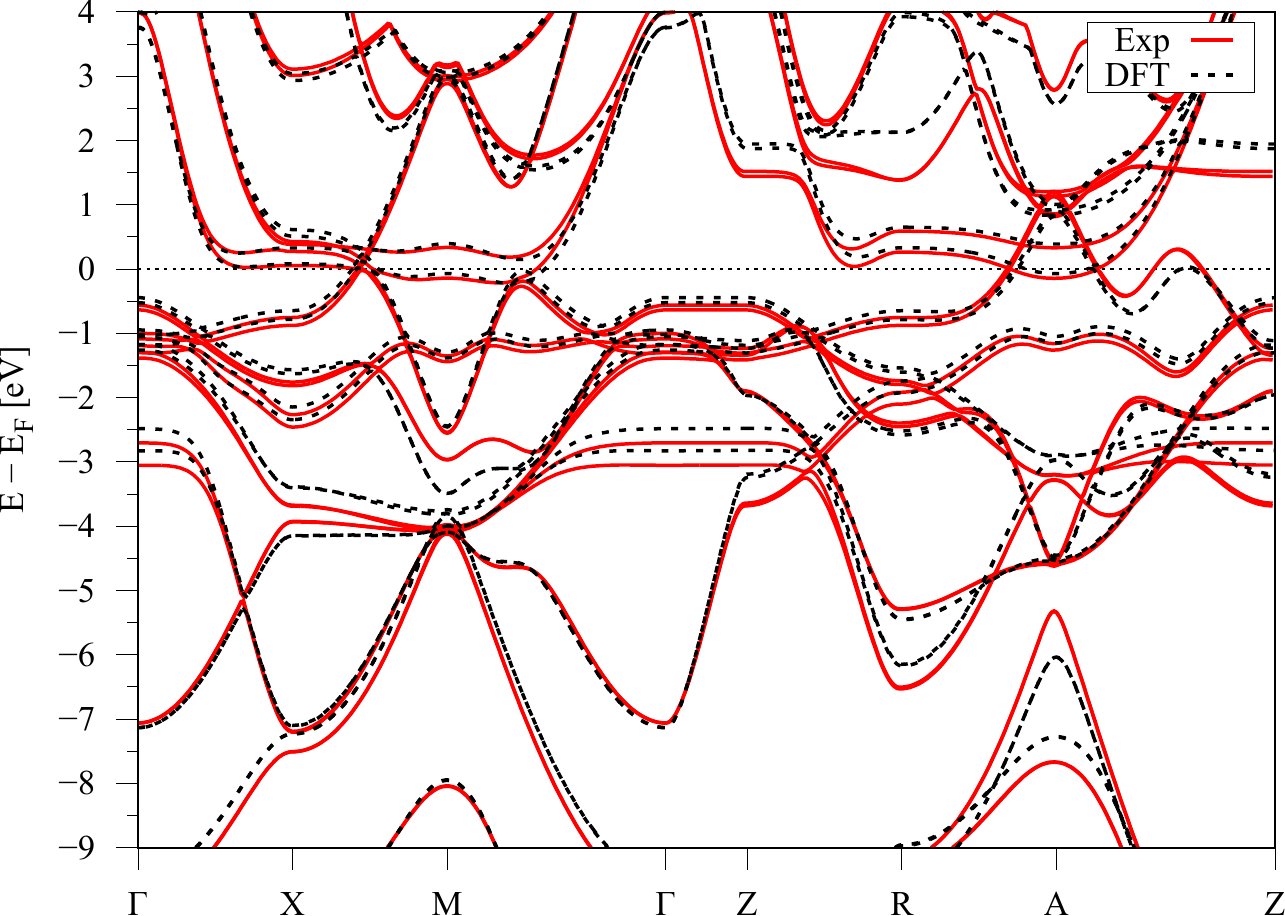}
    	\caption{Calculated band structure using DFT of FeGe$_2$ with SOC for the experimentally determined (solid red lines)
    	and for the fully relaxed (dashed black lines) lattice structure. }
    	\label{fig:bands_compare_EXP_and_DFT}
    \end{figure}

In Fig. \ref{fig:bands_compare_EXP_and_DFT}, the calculated band structures of FeGe$_2$ are compared for the cases of the
experimentally determined and the fully relaxed lattice structure. The lattice constants and magnetic moments have already been
discussed above in the geometry section. The overall band structures features are nearly the same.

\subsection{Magnetic Ground State}

   \begin{figure}[b]
    	\includegraphics[width=0.99\textwidth]{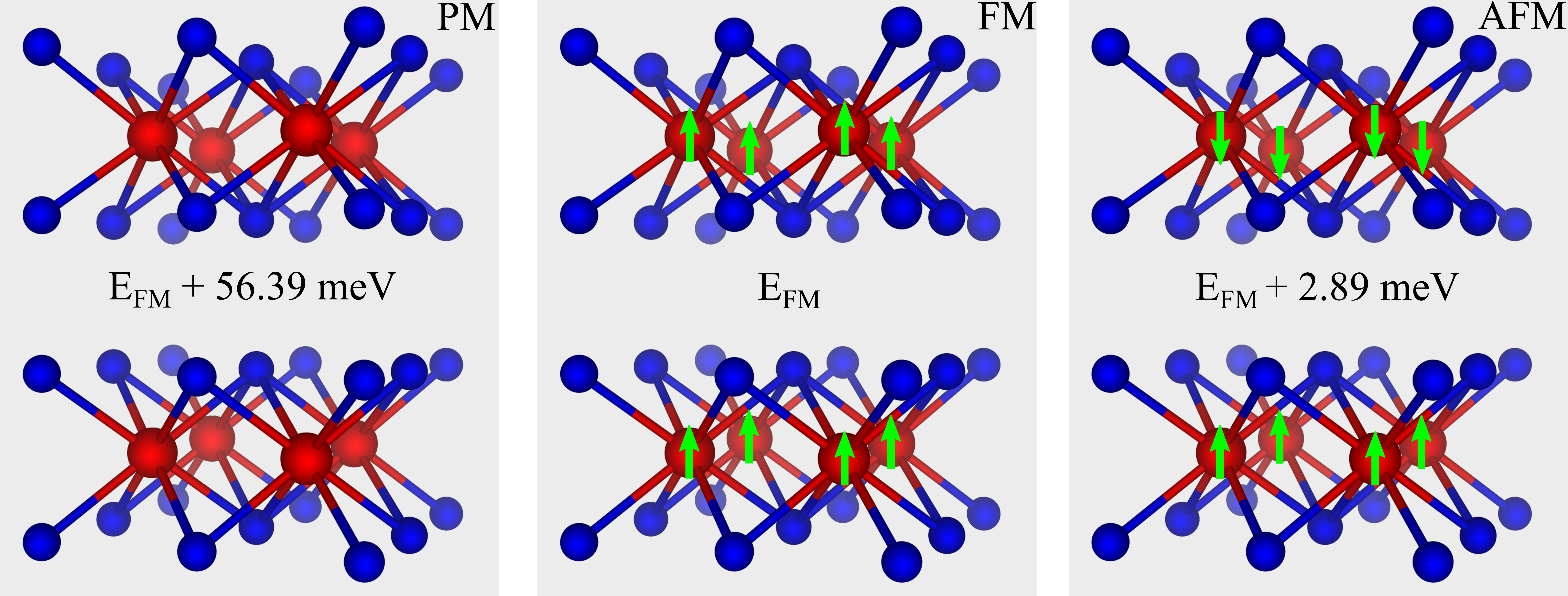}
    	\caption{Possible magnetic ground states of FeGe$_2$: paramagnetic (PM), ferromagnetic (FM),
    	and antiferromagnetic (AFM) state. Green arrows depict the magnetization of the Fe atoms.
    	The total energies are given with respect to the ground state energy of the FM state.  }
    	\label{fig:magnetic_states}
    \end{figure}
To determine the magnetic ground state of $\alpha$-FeGe$_2$, we consider a supercell of $2\times 2\times 2$ of FeGe$_2$ and
three different magnetic configurations as depicted in Fig. \ref{fig:magnetic_states}:
(i) paramagnetic (PM) phase, where all atoms are considered to be non-magnetic; (ii) ferromagnetic (FM) phase, in which all Fe atoms have aligned magnetizations along c-axis;
(iii) antiferromagnetic (AFM) phase, where individual FeGe$_2$ layers are ferromagnetically aligned.
From the total energy analysis of these cases, we find that the FM phase is the energetically most favorable ground state.
The PM phase is about 56~meV higher in energy (or about 7~meV per unit cell) compared to the FM phase, while the AFM phase is only about 2.9~meV higher in energy (or about 0.36~meV per unit cell).

To determine the ground state of the relaxed structure, we performed the same analysis and found that the FM phase is about
35~meV (19~meV) lower in energy than the PM (AFM) phase so that the FM phase corresponds again to the ground state.


%